\documentclass{jltp}

\usepackage{graphicx}

\title{Velocity of sound in a Bose-Einstein condensate in the presence
of an optical lattice and transverse confinement}

\author{M. Kr\"{a}mer$^{1,2}$, C. Menotti$^{1,2}$, and M. Modugno$^{2,3}$}

\address{$^{1}$Dipartimento di Fisica, Universit\`a di Trento, 
I-38050 Povo, Italy\\
 $^{2}$BEC-INFM Trento, I-38050 Povo, Italy\\
  $^{3}$LENS - Dipartimento di Fisica, Universit\`a di Firenze and INFM,\\
  I-50019 Sesto Fiorentino, Italy}

\runninghead{M. Kr\"{a}mer, C. Menotti, and M. Modugno}{Velocity of
sound in a Bose-Einstein condensate in an optical lattice}

\begin{document}

\maketitle

\begin{abstract} We study the effect of the transverse degrees of freedom
on the velocity of sound in a Bose-Einstein condensate immersed in a
one-dimensional optical lattice and radially confined by a harmonic trap. 
We compare the results of full three-dimensional calculations
with those of an effective 1D model based on the equation of state of the
condensate.  The perfect agreement between the two approaches is
demonstrated for several optical lattice depths and throughout the full 
crossover from the 1D mean-field to the Thomas Fermi regime in the radial 
direction.
\end{abstract}

Ever since the achievement of Bose-Einstein condensation (BEC) with atomic
vapors, an extensive activity in the study of the excitations of these
systems has been carried out.  A particular example is the exploration of
sound excitations in elongated condensates
\cite{andrews1997a,andrews1998a,zaremba1997a,kavoulakis1997a,stringari1998a}.
The propagation of sound in one-dimensional (1D) optical lattices has been
the subject of recent studies performed both in the linear
\cite{kraemer2002a,machholm2003a,smerzi2003a,kraemer2003a,taylor2003a,martikainen2004a}
and nonlinear regimes \cite{menotti2004a}.  To capture the main physics,
the system can be conveniently described by means of effective 1D
theories, which account for the transverse degrees of freedom through a
renormalization of the coupling constant. Still, it is important to check
the effect of the transverse degrees of freedom since they are known to
play a relevant role in some cases \cite{modugno2004a}.

In this paper, we show that the sound velocity in presence of a 1D lattice
and transverse harmonic confinement can be obtained by employing an
effective 1D approach, and demonstrate that the effect of the transverse
harmonic confinement is completely accounted for by the expression
\cite{machholm2003a,smerzi2003a,kraemer2003a,taylor2003a} 
\begin{eqnarray} 
c = \sqrt{\frac {N}{m^*}\frac{ \partial \mu(N)}{\partial N}} , 
\label{sound_vel} 
\end{eqnarray} 
where $N$ is the number of atoms per lattice well and $m^*$ is the
effective mass.  The effect of lattice and transverse confinement both
on the equation of state $\mu(N)$ and $m^*$ is discussed in detail.


Let us consider an axially symmetric condensate, which is radially
confined by the potential $V_{\rm ho}=m \omega_{\perp}^2 r^2/2$, and
immersed in a 1D periodic potential of the form $V_{\rm lat}(z)= s\;
E_R \;{\rm sin}^2\left( {\pi z / d} \right)$, where $d$ is the lattice
period, $E_R=q_B^2/2m$ the recoil energy, $q_B=\hbar\pi/d$ the
Bragg-momentum, and $s$ the lattice depth in units of the recoil
energy.  We consider any additional axial confinement to be 
negligible, so that the Gross-Pitaevskii (GP) equation takes the form
\begin{eqnarray}    
\left[-{\hbar^2 \over 2m}{\bf \nabla}^2 + V_{\rm ho}(r)
+V_{\rm lat}(z) +{g N }|\Psi(r,z)|^2 \right]\Psi(r,z)= {\mu }\Psi(r,z)\,,
\label{gpe-3d}    
\end{eqnarray}   
where $N$ is the number of atoms per lattice well and the condensate
wavefunction is normalized according to $\int d^2r\int_{-d/2}^{d/2}dz
|\Psi(r,z)|^2 = 1$.

In the linear regime, the small oscillations around the
condensate ground state can be described by expanding the order
parameter as $\Phi(r,z,t)=e^{-i\mu t/\hbar} [ \Psi(r,z) + \sum_{j\nu
q} u_{j\nu q} e^{-i \omega_{j\nu}(q) t} + v^*_{j\nu q} e^{i
\omega_{j\nu}(q) t} ]$, where the functions $u_{j\nu q}$ and $v_{j\nu
q}$ are the Bogoliubov quasiparticle amplitudes of quasimomentum $q$,
and quantum numbers $j,\nu$ representing the band index and the number
of radial nodes respectively.  In particular, the lowest energy 
solutions, with $j=1$ and $\nu=0$, are axial phonons that propagate
with the sound velocity $c$.  The latter is defined as the slope of
the lowest branch in the excitation spectrum $\omega_{j\nu}(q)$,
through the relation $c=\hbar \partial\omega_{10}(q)/\partial
q|_{q=0}$. Therefore it can be obtained by numerically solving the
spectrum of the full 3D system, as discussed in \cite{modugno2004a}.

In the following we will show that the system can be equivalently
described by an effective 1D approach.  The basic assumption we make
is that the condensate wavefunction can be factorised into a
longitudinal and a radial part $\Psi(r,z)=\varphi(r) \psi(z)$. The
physics underlying this assumption is that interactions mainly affect
the radial wavefunction $\varphi(r)$, while the axial wavefunction
$\psi(z)$ is that of a single particle in the periodic potential. This
factorization is exact for $s=0$, where the system is uniform in the
longitudinal direction.  In the case $s\neq 0$, the solution of the GP
equation for a transversally uniform system shows that for the
densities of many current experiments
\cite{denschlag,cataliotti,morsch1,greiner1} the equation of state and the
effective mass can be calculated neglecting the density dependence of
$\psi(z)$ \cite{kraemer2003a}. The remaining relevant density
dependence of these quantities arises from their explicit dependence
on the number of particles (see below). The range of densities for
which the factorization ansatz is correct becomes larger as the
lattice depth $s$ is increased.

Under the factorization assumption, and using the normalisation
condition for the wavefunction, one can write two effective GP
equations
\begin{eqnarray}    
&& \Bigg[-{\hbar^2 \over 2m}{\bf \nabla}_r^2 + V_{\rm ho}(r) + \left(
g \int_{-d/2}^{d/2} |\psi(z)|^4 dz \right) N |\varphi(r)|^2
\Bigg]\varphi(r)= \mu_\perp \varphi(r), \;\;\;\;
\label{gpe-r}  \\
&& \Bigg[ - {\hbar^2 \over 2m} \nabla_z^2 +V_{\rm lat}(z) +\left( g
\int |\varphi(r)|^4 d^2 r \right) N |\psi(z)|^2
\Bigg]\psi(z)= \mu_z \psi(z), \;\;\;\;
\label{gpe-z} 
\end{eqnarray}   
where $N$ is the number of particles in each well at equilibrium,
$\mu_z= \mu - \epsilon_{\rm ho}$ and $\mu_\perp=\mu-\epsilon_{\rm
lat}$, being $\epsilon_{\rm ho}=\int d^2 r\varphi^*(r) \left[-{\hbar^2
\over 2m}{\bf \nabla}_r^2 + V_{\rm ho}(r) \right]\varphi(r)$ and
$\epsilon_{\rm lat}$ the single particle ground state energy 
of the periodic potential.  

The effective mass is calculated from the lowest energy Bloch band

\begin{eqnarray}
\varepsilon (k)= \int_{-d/2}^{d/2}     
{\psi}^*_{k}(z) 
\left[ - {\hbar^2\over 2m } \nabla_z^2 
+   V_{\rm lat}(z)
+ {g\tilde{N}\over 2 a_{\perp}^2} |\psi_{k}(z)|^2 \right] 
\psi_{k}(z) dz,
\label{energy}
\end{eqnarray}
according to the general relation
$1/m^*=\partial^2\varepsilon/\partial k^2|_{k=0}$ \cite{kraemer2003a}.
In Eq.(\ref{energy}) the explicit dependence of the energy per
particle on density appears through the renormalized number of atoms
${\tilde N}=N a_{\perp}^2\int |\varphi(r)|^4 d^2 r $, where 
$a_{\perp}=\sqrt{\hbar/m\omega_{\perp}}$. Consistently
with the above discussion, the Bloch state $\psi_k(z)$ at
quasi-momentum $k$ is taken as the single particle solution of
(\ref{gpe-z}).  Then $\psi_k(z)$ is independent of the radial
confinement and the effect of the radial trapping on $m^*$ is captured
by its dependence on the renormalized number of atoms ${\tilde N}$.

Equation~(\ref{gpe-r}) describes a BEC uniform along $z$ and
harmonically confined along $r$ (infinite cylinder).  The effect
of the lattice is included in the effective coupling constant
\begin{eqnarray}
{\tilde g} = g d \int_{-d/2}^{d/2} |\psi(z)|^4 dz,
\label{gtilde}
\end{eqnarray}
where again $\psi(z)$ is the single particle ground state of
(\ref{gpe-z}).  Without radial confinement, in a system with 3D
average density $n$, the equation of state $\mu_{\perp}(n) = {\tilde
g} n$ is linear in the density, since ${\tilde g}$ depends strongly on
the optical lattice depth, but not on density.  As mentioned above,
this result is known to be correct in the presence of a lattice for
sufficiently small densities \cite{kraemer2003a}.

In order to discuss the effect of the transverse degrees of freedom,
we concentrate now on the radial dynamics. The presence of the lattice
is described by the effective coupling constant ${\tilde g}$, so that
the crossover from frozen radial dynamics (1D mean-field) to the
Thomas-Fermi (TF) regime, can be described in full analogy to the case
of the infinite cylinder \cite{menotti}.  Introducing the
dimensionless variable $\rho=r/a_{\perp}$ and defining
$f(\rho)=a_{\perp} \varphi(r)$, Eq.(\ref{gpe-r}) can be rewritten in
the dimensionless form

\begin{equation}
\left[-{1\over 2}{\bf \nabla}_{\rho}^2 +{1\over 2}\rho^2+4\pi {\tilde
a} n_{1D}|f(\rho)|^2\right] f(\rho) =
{\mu_\perp\over\hbar\omega_\perp}f(\rho)\,,
\label{gpe-1D}
\end{equation}
where $n_{1D}=N/d$ is the linear average density, ${\tilde a}$ is the
effective scattering length obtained from the relation ${\tilde g} = 4
\pi \hbar^2 {\tilde a}/m$, and $f$ is normalized according to $\int
d^2{\bf\rho} |f|^2=1$.  Notice that the solution of the above equation
depends only on the parameter ${\tilde a}n_{1D}$, yielding an exact
scaling behavior for the radial wavefunction as a function of the
transverse trapping.

The dependence of the chemical potential $\mu_\perp/\hbar\omega_\perp$
on ${\tilde a} n_{1D}$ determines the equation of state in the
presence of the radial trap.  For ${\tilde a} n_{1D}\gg 1$, the system
is in the TF regime and one finds
${\mu_\perp/\hbar\omega_\perp}=2({\tilde a} n_{1D})^{1/2}$.  In the
opposite limit ${\tilde a} n_{1D}\ll 1$, instead, the radial profile
is gaussian and the equation of state is given by
${\mu_\perp/\hbar\omega_\perp}=1+2{\tilde a} n_{1D}\,$.  Given the
solution $\mu_\perp/\hbar\omega_\perp$ as a function of the universal
parameter ${\tilde a} n_{1D}$, the velocity of sound can be calculated
according to Eq.(\ref{sound_vel}).
This procedure yields the result $c=\sqrt{gn(0)/2m^*}$ both in the TF
regime (${\tilde a} n_{1D}\gg 1$) and in the 1D mean-field limit
(${\tilde a} n_{1D}\ll 1$) \cite{taylor2003a}.  To find these limiting
results, we have used the fact that $n_{1D}=(4\pi^2\hbar^2
{\tilde a}/m^2\omega_\perp^2)n^2(0)$ in the TF regime and $n_{1D}=\pi
a_{\perp}^2 n(0)$ in the 1D mean-field limit.

\begin{figure}
\centerline{\includegraphics[height=2.5in]{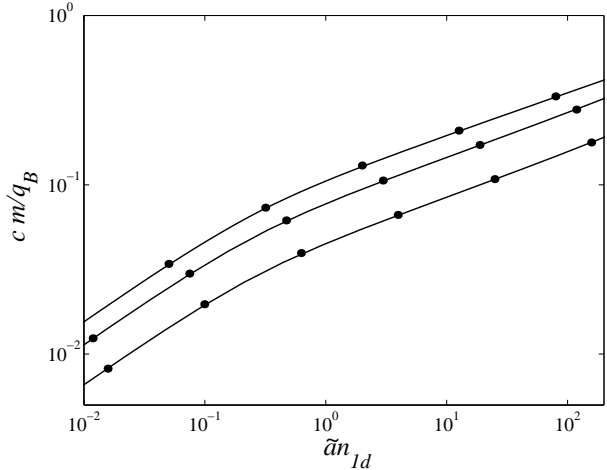}}
\caption{Velocity of sound as a function of ${\tilde a}n_{1D}$, for
three values of the lattice depth (from top to bottom $s=0,5,10$).
Results of full 3D numerical calculations (points) are compared with
the prediction derived from the 1D equation of state (lines). Note
that $\tilde{a}/a=1.49$ and $\tilde{a}/a=1.98$ for $s=5$ and $s=10$
respectively.}
\label{fig}
\end{figure}

The result for the sound velocity in the intermediate regimes can be
calculated numerically, just based on the knowledge of the following
quantities: (i) the renormalized scattering length ${\tilde a}$ for
any given value of $s$, obtained from a 1D calculation of the single
particle ground state solution in presence of the lattice; (ii) the
equation of state and ${\tilde N}$ for the infinite cylinder at any
${\tilde a}n_{1D}$; (iii) the effective mass $m^*$ at any given value
of $s$ for the renormalised number of atoms per well ${\tilde N}$,
obtained from Eq. (\ref{energy}) by a 1D calculation of the single
particle Bloch solutions.  This procedure leads to the results shown
in Fig.\ref{fig}, where the theoretical predictions are compared
with the results of the full 3D numerical calculations for several
values of the lattice depth.  The agreement is perfect in the full
crossover.  

Furthermore, it is interesting to note that by relating $n_{1d}$ to
the density at the center $n(0)$ one can show that the value
$c=\sqrt{{\tilde g}n(0)/2m^*}$ provides an estimate for the sound
velocity which is correct within $5\%$ throughout the whole crossover
from TF to 1D.

We also notice that in the TF regime, ${\tilde a}n_{1D}\gg 1$, the
sound velocity in presence of transverse harmonic confinement
corresponds to that of a transversally uniform condensate with the
average density ${\bar n}$, $c=\sqrt{{\tilde g}{\bar n}/2m^*}$.  In
fact, in this regime the radial profile is an inverted parabola,
leading to an average density ${\bar n} \equiv \int dxdy n(x,y)/\pi
d R^2$ equal to half of the central density $n(0)$
\cite{zaremba1997a,kavoulakis1997a,stringari1998a}.  This explanation
in terms of $\bar{n}$ works well in the TF regime, and to some
extent also in the 1D mean-field regime, where the definition for the
average density is replaced by ${\bar n} \equiv \int dxdy n^2(x,y) /
\int dxdy n(x,y)$ \cite{jackson1998}.  
However, since the concept of average density is
not well defined in general, it does not provide a general criterion
for the determination of the sound velocity in the intermediate
regimes.


In conclusion, we have demonstrated that the equation of state of an
infinite cylinder captures the effect of the transverse degrees of
freedom on the sound velocity in a condensate immersed in a 1D optical
lattice, upon a suitable renormalization
of the mass and of the coupling constant.  This result is obtained by
means of a factorization ansatz for the condensate wavefunction.  The
predictions of the factorized model yield perfect agreement with full
3D numerical calculations in a wide range of optical lattice depths
and in the full crossover from the 1D mean field regime to the TF
limit.  The factorization ansatz rely on the assumption that in
absence of radial trapping the equation of state is linear in the
number of particles.  This condition is always fulfilled at $s=0$ and
in the tight binding regime, while for intermediate $s$ it is
satisfied to a good approximation at experimentally relevant
densities.  In any case, Eq.(\ref{sound_vel}) is completely general,
and could be easily checked also beyond the limit of validity of the
factorization ansatz. In that case the determination of $\mu(N)$ would
require from the beginning a full 3D calculation.  Instead, when the
factorization assumption holds, the simple model presented in this
paper can be used to extract the correct value of the sound velocity.

\section*{ACKNOWLEDGMENTS}
This research is partially supported by the Mi\-ni\-ste\-ro
dell'Istru\-zio\-ne, dell'Uni\-ver\-si\-t\`a e del\-la Ri\-cer\-ca
(MIUR).  We thank Franco Dalfovo, Lev Pitaevskii, Augusto Smerzi and
Sandro Stringari for useful discussions.


\end{document}